\documentclass[aps,preprint,prd,showpacs,showkeys,nofootinbib,tightenlines]{revtex4-1}
\usepackage{graphicx}  
\usepackage{amsmath}
\usepackage{bm}  
\usepackage{ulem}
\usepackage{amssymb}
\usepackage{comment}
\usepackage{lineno}
\newcommand{\bea}{\begin{eqnarray}}
\newcommand{\eea}{\end{eqnarray}}
\newcommand{\beq}{\begin{equation}}
\newcommand{\eeq}{\end{equation}}
\newcommand{\bqa}{\begin{eqnarray}}
\newcommand{\eqa}{\end{eqnarray}}

\begin{document}

\title{
Production of $\bm{X(3872)}$ Accompanied by a Soft Pion\\
 at Hadron Colliders}

\author{Eric Braaten}
\email{braaten.1@osu.edu}
\affiliation{Department of Physics,
         The Ohio State University, Columbus, OH\ 43210, USA}

\author{Li-Ping He}
\email{he.1011@buckeyemail.osu.edu}
\affiliation{Department of Physics,
         The Ohio State University, Columbus, OH\ 43210, USA}

\author{Kevin Ingles}
\email{ingles.27@buckeyemail.osu.edu}
\affiliation{Department of Physics,
         The Ohio State University, Columbus, OH\ 43210, USA}

\date{\today}
%\date{November 2007}

\begin{abstract}
If the $X(3872)$ is a weakly bound charm-meson molecule, 
it can be produced by the creation of $D^{*0} \bar{D}^0$ or $D^{0} \bar{D}^{*0}$ at short distances
followed by the formation of  the bound state from the charm-meson pair. 
The $X$ can also be produced by the creation of $D^* \bar{D}^*$ at short distances
followed by  the rescattering of the charm-meson pair into $X \pi$.
At a high-energy hadron collider, the prompt cross section  from  this mechanism 
has a narrow peak in the $X \pi$ invariant mass distribution near the $D^* \bar D^*$ threshold
from a charm-meson triangle singularity.
An order-of-magnitude estimate of the ratio 
of the cross section for producing the peak in the $X \pi^\pm$ distribution 
to the cross section for producing $X$ without an accompanying pion 
suggests that the peak may be observable at the LHC.
\end{abstract}

\smallskip
\maketitle
%%%%%%%%%%%%%%%%%%%%%%%%%%%%%%%%%%%%%%%%%%
%%%%%%%%%%%%%%%%%%%%%%%%%%%%%%%%%%%%%%%%%%

\section{Introduction}
\label{sec:Introduction}

%\begin{linenumbers}

The discovery of a large number of exotic hadrons containing a heavy quark and its antiquark 
presents a major challenge to our understanding of QCD
\cite{Chen:2016qju,Hosaka:2016pey,Lebed:2016hpi,Esposito:2016noz,Guo:2017jvc,Ali:2017jda,Olsen:2017bmm,Karliner:2017qhf,Yuan:2018inv,Brambilla:2019esw}.
The first of these exotic hadrons to be discovered was the $X(3872)$ meson: it
was discovered in 2003 in exclusive decays of $B^\pm$ mesons into $K^\pm X$ 
through the decay of $X$ into $J/\psi\, \pi^+\pi^-$ \cite{Choi:2003ue}.
Its existence was quickly verified through inclusive production in $p \bar p$ collisions \cite{Acosta:2003zx}.
The $J^{PC}$ quantum numbers of $X$ were eventually determined to be $1^{++}$ \cite{Aaij:2013zoa}.
Its mass  is extremely close to the $D^{*0} \bar D^0$  threshold,
with the difference being only $0.01 \pm 0.18$~MeV \cite{Tanabashi:2018oca}.
This suggests that  $X$ is a weakly bound S-wave charm-meson molecule
with the flavor structure  \cite{Close:2003sg,Voloshin:2003nt}
%===============
\begin{equation}
\big| X(3872) \rangle = \frac{1}{\sqrt2}
\Big( \big| D^{*0} \bar D^0 \big\rangle +  \big| D^0 \bar D^{*0}  \big\rangle \Big).
\label{Xflavor}
\end{equation}
%===============
However, there are alternative models for the $X$
\cite{Chen:2016qju,Hosaka:2016pey,Lebed:2016hpi,Esposito:2016noz,Guo:2017jvc,Ali:2017jda,Olsen:2017bmm,Karliner:2017qhf,Yuan:2018inv,Brambilla:2019esw}.
The $X$ has been observed in many more decay modes than any of the other exotic heavy hadrons.
In addition to the discovery mode $J/\psi\, \pi^+\pi^-$, it has been observed in 
$J/\psi\, \pi^+\pi^-\pi^0$, $J/\psi\, \gamma$, $\psi(2S)\, \gamma$, $D^0 \bar D^0 \pi^0$, $D^0 \bar D^0 \gamma$, 
and most recently $\chi_{c1}\, \pi^0$ \cite{Ablikim:2019soz}.
The observation of $X$ in these 7 different decay modes has not 
proven to be effective in discriminating between these models.
There may be aspects of the production of $X$ that are more effective at discriminating
between models than the decays of $X$.

The $X$ can be produced by any reaction that can produce its constituents $D^{*0} \bar D^0$ and $D^0 \bar D^{*0}$.
In particular, it can be produced in high energy hadron collisions.
The inclusive production of $X$ in $p \bar p$ collisions has been studied at the Tevatron by the 
CDF \cite{Acosta:2003zx} and D0 \cite{Abazov:2004kp} collaborations.
The inclusive production of $X$ in $p p$ collisions has been studied at the 
Large Hadron Collider (LHC) by the LHCb \cite{Aaij:2011sn},
CMS \cite{Chatrchyan:2013cld}, and ATLAS \cite{Aaboud:2016vzw} collaborations.
At a high energy hadron collider,  $X$ is produced by the weak decays of bottom hadrons,
and it is also  produced promptly through QCD mechanisms that create charm quarks and antiquarks. 
The substantial prompt production rate of $X$ at hadron colliders has often been used as an
argument against its identification as a weakly bound charm-meson molecule.  
This argument is based on an upper bound on the cross section for producing $X$
in terms of the cross section for producing the charm-meson pair $D^{*0} \bar D^0$  
integrated up to a maximum relative momentum $k_\mathrm{max}$ \cite{Bignamini:2009sk}.
The estimate for $k_\mathrm{max}$  in Ref.~\cite{Bignamini:2009sk} was 
approximately the binding momentum $\gamma_X$ of the $X$.
In Ref.~\cite{Artoisenet:2009wk}, it was pointed out that the derivation of the upper bound 
in Ref.~\cite{Bignamini:2009sk} requires $k_\mathrm{max}$ to be of order the pion mass $m_\pi$ instead of $\gamma_X$.
In Ref.~\cite{Braaten:2018eov}, we used the methods of  Ref.~\cite{Artoisenet:2009wk} 
to derive an equality for the $X$ cross section.  
The resulting estimates for the prompt cross sections for $X$ at the Tevatron and at the LHC 
are compatible with experimental lower bounds on the cross sections.  

The production of $X$ can also proceed through the creation of a  pair  
of spin-1 charm mesons $D^*\bar D^*$ at short distances followed by their rescattering into $X\pi$.
In Ref.~\cite{Braaten:2019yua}, we studied the decays of $B$ mesons into $KX\pi$
from the decay at short distances into $K$ plus a $D^*\bar D^*$ pair
followed by the rescattering of the charm-meson pair into $X\pi$.
We used a previous isospin analysis of the decays $B \to K D^{(*)} \bar D^{(*)}$ \cite{Poireau:2011gv}
to estimate the short-distance amplitudes for creating the $D^*\bar D^*$ pair.
We used XEFT to calculate the amplitudes for the rescattering of  $D^*\bar D^*$ 
to $X \pi$ with small relative momentum.
The differential decay rate has a narrow peak in the $X \pi$ invariant mass near the $D^*\bar D^*$ threshold
from a charm-meson triangle singularity.
We obtained an estimate of the branching fraction into the peak from the charm-meson triangle singularity
in the decays $B^0 \to K^+X\pi^-$  and $B^+ \to K^0X\pi^+$ \cite{Braaten:2019yua}. 

In this paper, we study the inclusive prompt production
of $X\pi$ from rescattering of $D^*\bar D^*$ in high energy hadron collisions.
The outline of this paper is as follows.
In Section~\ref{sec:Molecule}, we describe some universal aspects of weakly bound S-wave molecules
as well as an effective field theory XEFT for charm mesons and pions. 
In the subsequent sections, we apply XEFT to various cross sections at a high energy hadron collider.
We consider the production of $X$ in Section~\ref{sec:Equality}
and the production of a pair of spin-1 charm mesons in Section~\ref{sec:D*D*}.
In  Section~\ref{sec:Xsoftpi}, we
calculate the cross section for producing $X\pi$ with invariant mass near the $D^* \bar D$ threshold.
We summarize our results and discuss their implications in  Section~\ref{sec:Summary}.

%\newpage 

%%%%%%%%%%%%%%%%%%%%%%%%%%%%%%%%%%%%%%%%%%
\section{Weakly Bound S-wave  Molecule and XEFT}
\label{sec:Molecule}
%%%%%%%%%%%%%%%%%%%%%%%%%%%%%%%%%%%%%%%%%%

If short-range interactions produce an S-wave bound state extremely close to a scattering threshold,
the few-body physics has universal aspects that are determined by the {\it binding momentum} $\gamma_X$ 
of the bound state \cite{Braaten:2004rn}. 
The binding energy is $\gamma_X^2/(2\mu)$, where $\mu$ is the reduced mass of the constituents.
The momentum-space wavefunction in the region of relative momentum $k$ below the inverse range 
has the universal form
%===============
\begin{equation}
\psi_X(k) = \frac{ \sqrt{8 \pi \gamma_X}}{k^2 + \gamma_X^2}.
\label{psiX-k}
\end{equation}
%===============

The low-energy scattering of the constituents also has universal aspects determined by $\gamma_X$
through a simple function of the complex energy $E$ relative to the scattering threshold:
%===============
\begin{equation}
f_X(E) = \frac{1}{-\gamma_X+\sqrt{-2\mu E }}.
\label{f-E}
\end{equation}
%===============
The universal elastic scattering amplitude  in the region of relative momentum $k$ below the inverse range 
is obtained by evaluating this function at energy $E=k^2/2\mu + i \epsilon$.
The function $f_X(E)$ also gives the energy distribution from creation of the constituents at short distances. 
 By the optical theorem, the distribution in the energy $E$ 
below the energy scale set by  the range  is proportional to the imaginary part of $f_X(E)$: 
%===============
\begin{equation}
\mathrm{Im}[f_X(E+ i \epsilon)] = \frac{\pi \gamma_X}{\mu} \delta(E + \gamma_X^2/2\mu)
+ \frac{\sqrt{2\mu E}}{\gamma_X^2+2\mu E} \theta(E).
\label{Imf-E}
\end{equation}
%===============
There is a delta-function term at a negative energy  from the production of the weakly bound molecule
and a theta-function term  with positive energy from the  production of the constituents of the molecule.

If the $X(3872)$ is a weakly bound charm-meson molecule, its constituents are
the superposition of charm mesons in Eq.~\eqref{Xflavor}.
We denote the masses of the spin-0 charm mesons  $D^0$ and $D^+$ by $M_0$ and $M_1$,
the masses of the spin-1 charm mesons  $D^{*0}$ and $D^{*+}$ by $M_{*0}$ and $M_{*1}$
and the masses of the pions  $\pi^0$ and $\pi^+$ by $m_0$ and $m_1$
(or collectively by $m_\pi$).
The range of the interactions between the charm mesons is $1/m_\pi$.
The corresponding energy scale $m_\pi^2/2\mu$  is about 10~MeV.
This is comparable to the energy of the $D^{*+} D^-$ scattering threshold,
which is 8.2~MeV above the  $D^{*0} \bar D^0$  scattering threshold.
The present value of the difference $E_X$ between the mass of the $X$ 
and the energy of the $D^{*0} \bar D^0$ scattering threshold is \cite{Tanabashi:2018oca}
%===============
\begin{equation}
E_X \equiv M_X - (M_{*0}+M_0) =( +0.01 \pm 0.18)~\mathrm{MeV}.
\label{eq:deltaMX}
\end{equation}
%===============
The central value in Eq.~\eqref{eq:deltaMX} corresponds to a charm-meson pair above the scattering threshold.
The value lower by $1\sigma$ corresponds to a bound state with binding energy $|E_X| =0.17$~MeV
and binding momentum $\gamma_X=18$~MeV.

The universal results for the  wavefunction
for a near-threshold S-wave bound state in Eq.~\eqref{psiX-k} 
and the  scattering amplitude for its constituents in Eq.~\eqref{f-E} can be derived from 
a zero-range effective field theory (ZREFT) with a single scattering channel  \cite{Braaten:2004rn}.
In the case of $X$, the constituents are the 
neutral charm mesons $D^{*0}$, $\bar D^{*0}$, $D^0$, and $\bar D^0$ \cite{Braaten:2005jj}.
This ZREFT describes explicitly the  $D^{*0} \bar D^0$ and  $D^{0} \bar D^{*0}$ 
components of the $X$ with energies sufficiently close to their scattering threshold.
Its range of validity extends at most up to the $D^{*+} D^-$ scattering threshold,
which is higher in energy by 8.2 MeV and corresponds to a relative momentum for  $D^{*0} \bar D^0$ of 126 MeV.
This effective field theory does not describe explicitly the $D^{0} \bar D^0 \pi^0$ component of the $X$, 
which can arise from the decays $D^{*0} \to D^0 \pi^0$ or $\bar D^{*0} \to \bar D^0 \pi^0$. 

Fleming, Kusunoki, Mehen and van Kolck developed an effective field theory for $X$
called {\it XEFT} that has a much greater range of validity,
because it describes pion interactions explicitly \cite{Fleming:2007rp}.
It is an effective field theory for neutral and charged charm mesons $D^*$, $\bar D^*$, $D$, and $\bar D$
and for neutral and charged pions $\pi$.
XEFT describes explicitly the  $D^* \bar D$, $D \bar D^{*}$, and $\bar D D \pi$ components of the $X$.
The region of validity of the original formulation of XEFT extends to about the minimum energy  required to produce 
a $\rho$ meson.
For a charm meson pair,  the region of validity is below a relative momentum of about 1000~MeV.
For a charm meson pair plus a pion, the region of validity of XEFT is also  limited by the nonrelativistic approximation
for the pion: the relative momentum of the pion must be less than about $m_\pi \approx 140$~MeV.
We refer to a  pion with relative momentum  of order $m_\pi$ or smaller as a {\it soft pion}.

A Galilean-invariant formulation of XEFT that exploits the 
approximate conservation of mass in the transitions $D^* \leftrightarrow D \pi$
was developed in Ref.~\cite{Braaten:2015tga}.
In  Galilean-invariant XEFT, the spin-0 charm mesons $D^0$ and $D^+$ have the same kinetic mass $M_0$,
the spin-1 charm mesons $D^{*0}$ and $D^{*+}$ have the same kinetic mass $M_0+m_0$,
and the pions $\pi^0$ and $\pi^+$  have the same kinetic mass $m_0$.
The difference between the physical  mass and the kinetic mass of a particle
is taken into account through its rest energy.
Galilean invariance simplifies the utraviolet divergences of XEFT.
The pion number defined by the sum of the numbers of $D^*$, $\bar D^*$, and $\pi$ mesons
is conserved in Galilean-invariant XEFT.
The region of validity of Galilean-invariant XEFT extends  up to about the minimum energy required 
to produce an additional pion, which is above the  $D^* \bar D$ threshold by about 140~MeV.
For a charm-meson pair, the region of validity extends to a relative momentum of about 500~MeV.

In Ref.~\cite{Braaten:2010mg}, Braaten, Hammer, and Mehen pointed out that XEFT 
could also be applied to sectors with pion number larger than 1.
In particular, it can be applied to the sector with pion number 2,
which consists of $D^* \bar D^*$, $D \bar D^* \pi$, $D^* \bar D \pi$, $D \bar D \pi \pi$, and $X\pi$.
The cross sections for $D^* \bar D^* \to D^* \bar D^*$ and $D^* \bar D^* \to X \pi$ at small kinetic energies
were calculated in Ref.~\cite{Braaten:2010mg}.
If a high energy process can create $D^* \bar D^*$  at short distances,
XEFT can describe their subsequent rescattering into $X$ plus a soft pion.
In Ref.~\cite{Braaten:2019yua}, we applied XEFT to exclusive decays of $B$ mesons into $K X \pi$. 
In sections~\ref{sec:D*D*} and \ref{sec:Xsoftpi} of this paper,
we apply XEFT to the inclusive prompt production of $D^* \bar D^*$
and $X \pi$ with small relative momentum in high energy hadron collisions.

%\newpage 

%%%%%%%%%%%%%%%%%%%%%%%%%%%%%%%%%%%%%%%%%%
\section{Production of $\bm{X}$}
\label{sec:Equality}
%%%%%%%%%%%%%%%%%%%%%%%%%%%%%%%%%%%%%%%%%%

If  $X(3872)$ is a weakly bound charm-meson molecule with the flavor structure in Eq.~\eqref{Xflavor},
the cross section for producing $X$ can  be expressed in terms of the amplitudes
for producing $D^{*0} \bar D^0$ and $D^0 \bar D^{*0}$ \cite{Bignamini:2009sk}.
The inclusive prompt cross sections for  producing $D^{*0} \bar D^0$ 
with small relative momentum $\bm{k}$ in the charm-meson-pair rest frame
and  for  producing $X$ can be expressed as
%===============
\begin{subequations}
\begin{eqnarray}
d\sigma[D^{*0} \bar D^0] &=&
\frac{1}{\mathrm{flux}} \sum_y \int d\Phi_{(D^*\bar D)+y}
\Big|  \mathcal{A}_{D^{*0} \bar D^0+y}(\bm{k}) \Big|^2 
\frac{d^3k}{(2 \pi)^32 \mu},
\label{sigmaDstarD}
\\
%d\sigma[D^0 \bar D^{*0}] &=&
%\frac{1}{\mathrm{flux}} \sum_y \int d\Phi_{(D^*\bar D)+y}
%\Big|  \mathcal{A}_{D^{0} \bar D^{*0}+y}(\bm{k}) \Big|^2 
 %\frac{d^3k}{(2 \pi)^32 \mu},
%\label{sigmaDDstar}
%\\
d\sigma[X(3872)] &=&
\frac{1}{\mathrm{flux}} \sum_y \int d\Phi_{(D^*\bar D)+y}
\left|  \int \!\!\frac{d^3k}{(2 \pi)^3}  \psi_X(k) 
\frac{ \mathcal{A}_{D^{*0} \bar D^0+y}(\bm{k}) +  \mathcal{A}_{D^{0} \bar D^{*0}+y}(\bm{k})}{\sqrt2}\right|^2 
\frac{1}{2\mu},
\nonumber\\
\label{sigmaX}
\end{eqnarray}
\label{sigmaDstarD,DDstar,X}%
\end{subequations}
%===============
where $\mu$ is the reduced mass of $D^{*0} \bar D^0$.
The sums over $y$ are over all the additional particles that can be produced.
The cross section for $D^0 \bar D^{*0}$ is the same as in Eq.~\eqref{sigmaDstarD}
with $\mathcal{A}_{D^{*0} \bar D^0+y}(\bm{k})$ replaced by $\mathcal{A}_{D^{0} \bar D^{*0}+y}(\bm{k})$.
The charge-conjugation-even superposition of those amplitudes appears in the cross section for $X$ in Eq.~\eqref{sigmaX}.
The momentum-space wavefunction for the $X$ in Eq.~\eqref{sigmaX} is normalized so 
$ \int (d^3k/(2 \pi)^3)\, |\psi_X(k)|^2 = 1$.
The differential phase space $d\Phi_{(D^*\bar D)+y}$  is that for a composite particle 
 denoted by $(D^*\bar D)$ with mass $M_{*0} + M_0$ plus the additional particles $y$.  
 Factors of 3 from the sums over the spin states of $D^{*0}$ or $\bar D^{*0}$ or $X$
are absorbed into the amplitudes $ \mathcal{A}$.
The phase-space integrals in Eqs.~\eqref{sigmaDstarD,DDstar,X} are over the 3-momenta of the additional particles $y$,
 but the cross sections remain differential in the 3-momentum $\bm{P}$ of $(D^*\bar D)$.
 Thus the $D^{*0} \bar D^0$ cross section in Eq.~\eqref{sigmaDstarD} 
is differential in both $\bm{P}$ and $\bm{k}$, while the $X$ cross section in Eq.~\eqref{sigmaX}
is differential only in  $\bm{P}$.

Expressions for the cross sections in Eqs.~\eqref{sigmaDstarD,DDstar,X} that take into account 
the $X$ resonance were presented in Ref.~\cite{Artoisenet:2009wk}.
The cross sections in Eqs.~\eqref{sigmaDstarD,DDstar,X} were expressed in factored forms,
with long-distance factors that involve the binding momentum $\gamma_X$ and with 
short-distance factors that involve only momentum scales of order $m_\pi$ or larger.
The amplitudes for producing  $D^{*0} \bar D^0+y$ in Eq.~\eqref{sigmaDstarD} can be decomposed into 
charge-conjugation-even ($C=+$) and charge-conjugation-odd ($C=-$) components.
The $C=+$ component  is enhanced by the $X$ resonance.
If the nonresonant $C=-$ component is neglected, the amplitude for producing  $D^{*0} \bar D^0+y$
can be expressed as a product of the $C=+$ component of a short-distance
amplitude and a resonance factor that depends on $\gamma_X$:
%===============
\begin{equation}
 \mathcal{A}_{D^{*0} \bar D^0+y}(\bm{k})  = \frac{1}{\sqrt2} \left(
\frac{\mathcal{A}^{\mathrm{s.d.}}_{D^{*0} \bar D^0+y} + \mathcal{A}^{\mathrm{s.d.}}_{D^0 \bar D^{*0}+y}}{\sqrt2} \right)
\, \frac{\Lambda}{-\gamma_X-ik}.
\label{amp-DstarDbar}
\end{equation}
%===============
The expression for the corresponding amplitude $\mathcal{A}_{D^0 \bar D^{*0}+y}(\bm{k})$ is identical.
The short-distance amplitudes $\mathcal{A}^{\mathrm{s.d.}}_{D^{*0} \bar D^0+y}$
and $\mathcal{A}^{\mathrm{s.d.}}_{D^0 \bar D^{*0}+y}$
are independent of the momentum if $\bm{k}$ is small compared to  $m_\pi$.
The constant $\Lambda$ in the numerator of the resonance factor should be of order $m_\pi$.
The only dependence on the small momentum $\gamma_X$ is in the denominator of the resonance factor.
Since $\Lambda \gg \gamma_X$, the absolute value of the resonance factor is approximately 1 at $k = \Lambda$,
so $\Lambda$  can be interpreted as the momentum scale 
where  the amplitude becomes comparable in magnitude to the amplitude in the absence of the resonance.

The factorization formula for the $D^{*0} \bar D^0$ cross section can be obtained simply 
by inserting the amplitude in Eq.~\eqref{amp-DstarDbar} into Eq.~\eqref{sigmaDstarD}.
The factorization formula for the $X$ cross section  cannot be obtained so simply.
If the universal wavefunction in Eq.~\eqref{psiX-k} is inserted into Eq.~\eqref{sigmaX},
 the momentum integral is logarithmically ultraviolet divergent.
The factorization formula for the $X$ cross section can be obtained instead by requiring the sum of the cross sections 
for producing $X$ and the cross sections for producing $D^{*0} \bar D^0$ and $D^0 \bar D^{*0}$
integrated over $\bm{k}$ to be consistent with the optical theorem in Eq.~\eqref{Imf-E}.
The resulting factorization formulas for the inclusive prompt cross sections are
%===============
\begin{subequations}
\begin{eqnarray}
d\sigma[D^{*0} \bar D^0] &=&
\frac{1}{\mathrm{flux}} \sum_y \int d\Phi_{(D^*\bar D)+y}
\left| \mathcal{A}^{\mathrm{s.d.}}_{D^{*0} \bar D^0+y} + \mathcal{A}^{\mathrm{s.d.}}_{D^0 \bar D^{*0}+y} \right|^2 
\frac{\Lambda^2}{\gamma_X^2+k^2} \frac{d^3k}{(2 \pi)^3 8 \mu},
\label{factor-DstarDbar}
\\
%d\sigma[D^0 \bar D^{*0}] &=&
%\frac{1}{\mathrm{flux}} \sum_y \int d\Phi_{(D^*\bar D)+y}
%\left| \mathcal{A}^{\mathrm{s.d.}}_{D^{*0} \bar D^0+y} + \mathcal{A}^{\mathrm{s.d.}}_{D^0 \bar D^{*0}+y} \right|^2 
%\frac{\Lambda^2}{\gamma_X^2+k^2}  \frac{d^3k}{(2 \pi)^3 8 \mu},
%\label{factor-DDbarstar}
%\\
d\sigma[X(3872)] &=&
\frac{1}{\mathrm{flux}} \sum_y \int d\Phi_{(D^*\bar D)+y}
\left| \mathcal{A}^{\mathrm{s.d.}}_{D^{*0} \bar D^0+y} + \mathcal{A}^{\mathrm{s.d.}}_{D^0 \bar D^{*0}+y} \right|^2 
\frac{\Lambda^2\gamma_X}{8\pi\mu}.
\label{factor-X}
\end{eqnarray}
\label{factor-DstarD,X}%
\end{subequations}
%===============
The differential cross section for $D^{*0} \bar D^0$ in Eq.~\eqref{factor-DstarDbar} 
should be a good approximation up to relative momentum $k$ of about $\Lambda$.

In the expression for the $X$ cross section in Eq.~\eqref{factor-X}, there are  interference terms 
between the short-distance amplitudes for producing $D^{*0} \bar D^0+y$ and $D^0 \bar D^{*0}+y$.
The interference terms are positive for some sets of additional final-state particles $y$ and negative for others.
In high-energy hadron collisions, there are dozens or even hundreds of additional particles.
The sum over the many additional particles $y$ gives cancellations that suppress the interference terms. 
The $X$ cross section in Eq.~\eqref{sigmaX} then reduces to the sum of a $D^{*0} \bar D^0$ term 
and a $D^0 \bar D^{*0}$ term.  At large transverse momentum, the hadronization of a $c \bar c$ pair 
is equally likely to produce $D^{*0} \bar D^0$ and $D^0 \bar D^{*0}$,
because the probability of a light quark or antiquark from  a colliding hadron
to become a constituent of one of the charm mesons is very small.
The $D^{*0} \bar D^0$ term and the $D^0 \bar D^{*0}$ term should therefore be equal, 
and the expression for the $X$ cross section  reduces to 
%===============
\begin{equation}
d\sigma[X(3872)] =
\frac{1}{\mathrm{flux}} \sum_y \int d\Phi_{(D^*\bar D)+y}
\left| \mathcal{A}^{\mathrm{s.d.}}_{D^{*0} \bar D^0+y}\right|^2 
\frac{\Lambda^2\gamma_X}{4\pi\mu}.
\label{factor-X-interference}
\end{equation}
%===============

%\newpage 

%%%%%%%%%%%%%%%%%%%%%%%%%%%%%%%%%%%%%%%%%%
\section{Production of a Spin-1 Charm-Meson Pair}
\label{sec:D*D*}
%%%%%%%%%%%%%%%%%%%%%%%%%%%%%%%%%%%%%%%%%%

%%%%%%%%%%%%%%%%%%%%%%%%%%%%%%%%%%%%%%%%%%%%%%%%
\begin{figure}[t]
\includegraphics*[width=0.3\linewidth]{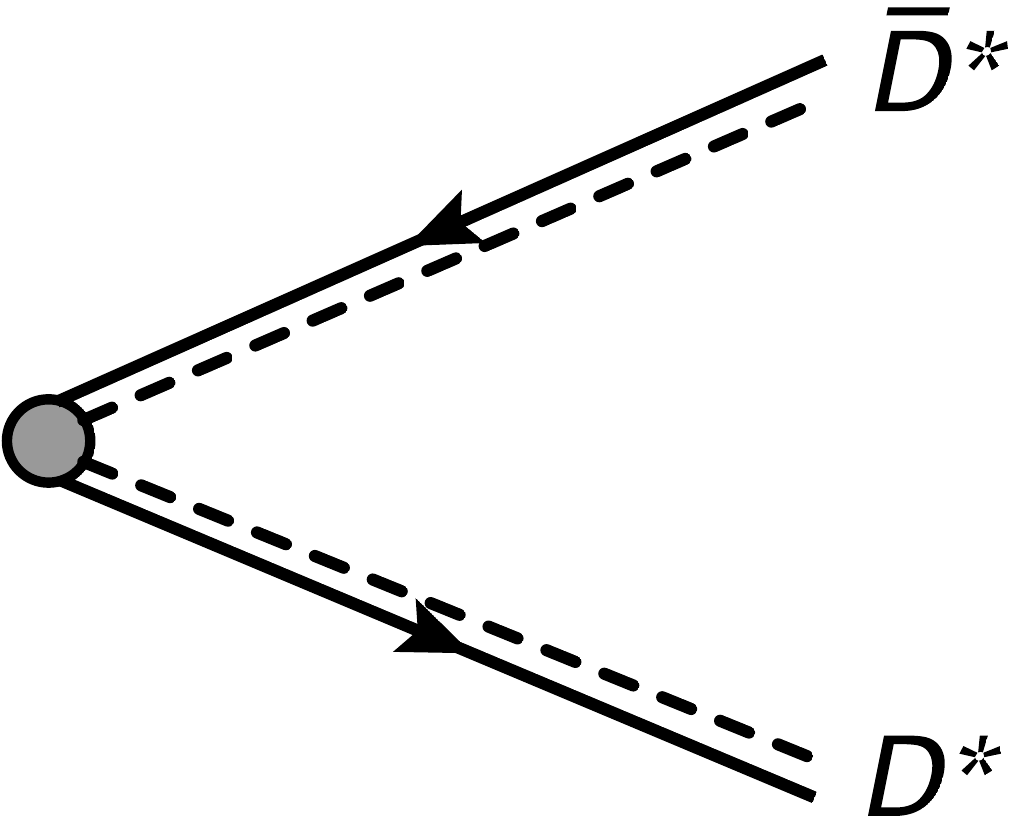} 
\caption{
Feynman diagram in XEFT for production of $D^* \bar D^*$ from their creation at a point.
The $D^*$ and $\bar D^*$ are represented by double lines consisting of  a dashed line
and a solid  line with an arrow. 
}
\label{fig:D*D*}
%\vspace*{0.0cm}
\end{figure}
%%%%%%%%%%%%%%%%%%%%%%%%%%%%%%%%%%%%%%%%%%%%%%

XEFT can be applied to the production of $D^* \bar D^*$ through a short-distance process 
that creates the pair of charm mesons.  
A short-distance process is one in which all the other particles in the reaction 
have momenta in the $D^* \bar D^*$ center-of-momentum (CM) frame that are much larger than $m_\pi$.
As far as the $D^*$ and $\bar D^*$ are concerned, the process can be described 
as a point interaction that creates $D^*$ and $\bar D^*$. 
The amplitude for producing $D^* \bar D^*$ can be represented in XEFT by  
the Feynman diagram in Fig.~\ref{fig:D*D*} with a vertex from which the $D^*$ and $\bar D^*$ emerge.
We denote the vertex factor for the reaction that creates $D^{*0} \bar D^{*0}$ at a point and produces the set of additional particles $y$ by $i \mathcal{A}^{ij}_{D^{*0} \bar D^{*0}+y}$, 
where $i$ and $j$ are the spin indices of the $D^{*0}$ and $ \bar D^{*0}$.
The matrix element for the reaction that produces $D^{*0} \bar D^{*0}$ with polarization vectors $\bm{\varepsilon}$ 
and $\bm{\bar{ \varepsilon}}$ is $\mathcal{A}^{ij}_{D^{*0} \bar D^{*0}+y} \varepsilon^i\bar \varepsilon^j $.

To obtain the differential cross section for producing $D^{*0} \bar D^{*0}$, the amplitude 
$\mathcal{A}^{ij}_{D^{*0} \bar D^{*0}+y}$ must be multiplied by its 
complex conjugate $(\mathcal{A}^{kl}_{D^{*0} \bar D^{*0}+y})^*$.
Their product must be summed over the additional particles $y$
(which includes sums over their spin states and integrals over their momenta),
summed over the spin states of $D^{*0}$ and $\bar D^{*0}$, 
integrated over the phase space of $D^{*0}$ and $\bar D^{*0}$, and divided by the flux factor.
The amplitude $\mathcal{A}^{ij}_{D^{*} \bar D^{*}+y}$ is a Cartesian tensor in the CM frame of $D^{*} \bar D^{*}$
with vector indices $ij$.  It must be a linear combination of $\delta^{ij}$,
terms of the form $\epsilon^{ijk} \hat P_m^k$ and $\hat P_m^i \hat P_n^j$
that involve unit vectors in the directions of the large momenta
of some of the additional particles $y$, and terms that involve polarization vectors and tensors
of the additional particles.  For each set of additional particles $y$,
the product of $\mathcal{A}^{ij}_{D^{*}\bar D^{*}+y}$ and $(\mathcal{A}^{kl}_{D^{*} \bar D^{*}+y})^*$
summed over their spins and integrated over their momenta
is a 4-index Cartesian tensor that defines a density matrix  in the spin indices $i j$ of the amplitude
and the spin indices $kl$ of the complex conjugate amplitude.
This tensor must be a linear combination of $\delta^{ij} \delta^{kl}$, $\delta^{ik} \delta^{jl}$, and $\delta^{il} \delta^{jk}$. 
The term $\delta^{ik} \delta^{jl}$ is diagonal in the spin indices $i k$ of the charm meson
and in the spin indices $jl$ of the anti-charm meson.
The terms $\delta^{ij} \delta^{kl}$ and  $\delta^{il} \delta^{jk}$ take into account entanglement between the spins.
Their coefficients will be positive for some sets of the additional particles $y$ and negative for other sets.
In the case of inclusive prompt production at the Tevatron or the LHC, cancellations from the sum 
over the many additional particles will suppress the $\delta^{ij} \delta^{kl}$ and  $\delta^{il} \delta^{jk}$ terms,
leaving a density matrix proportional to  $\delta^{ik} \delta^{jl}$.
The  weighted average from the sum over the additional particles $y$ therefore has the form
%===============
\begin{equation}
\left\langle  \mathcal{A}^{ij}_{D^{*0} \bar D^{*0}+y}\big(\mathcal{A}^{kl}_{D^{*0} \bar D^{*0}+y}\big)^*   \right\rangle
=  \frac19 \left\langle \big|  \mathcal{A}_{D^{*0} \bar D^{*0}+y}\big|^2 \right\rangle \delta^{ik} \delta^{jl} .
\label{AASS-D*Dbar*}
\end{equation}
%===============
After multiplying by the polarization vectors, the sum over the spin states of $D^{*0}$ and $\bar D^{*0}$ gives
%===============
\begin{equation}
\sum_\mathrm{spins}\left\langle\mathcal{A}^{ij}_{D^{*0} \bar D^{*0}+y} \big(\mathcal{A}^{kl}_{D^{*0} \bar D^{*0}+y}\big)^*\right\rangle
(\varepsilon^i \bar \varepsilon^j) (\varepsilon^k \bar \varepsilon^l )^*
=  \left\langle \big|  \mathcal{A}_{D^{*0} \bar D^{*0}+y}\big|^2 \right\rangle.
\label{sumAASS-D*0Dbar*0}
\end{equation}
%===============
The prefactor 1/9 in Eq.~\eqref{AASS-D*Dbar*} was chosen so the prefactor in Eq.~\eqref{sumAASS-D*0Dbar*0}
would be 1.

The differential cross section for producing $D^{*0} \bar D^{*0}$
with large total momentum and small relative momentum $\bm{k}$ 
and with polarization vectors $\bm{\varepsilon}$ and $\bm{\bar{ \varepsilon}}$ can be expressed as
%===============
\begin{eqnarray}
d\sigma[D^{*0} \bar D^{*0}] =
\frac{1}{\mathrm{flux}} \sum_y \int d\Phi_{(D^*\bar D^*)+y}
\Big| \mathcal{A}_{D^{*0} \bar D^{*0} +y}  \Big|^2  \frac{d^3k}{(2 \pi)^3M_{*0}}.
\label{factor-DstarDstar}
\end{eqnarray}
%===============
The differential phase space $d\Phi_{(D^*\bar D^*)+y}$ is that for a composite particle denoted by $(D^*\bar D^*)$ 
with mass $2M_{*0}$  and the additional particles $y$.
The integrals are over the 3-momenta of the additional particles but not over the 3-momentum $\bm{P}$ of $(D^*\bar D^*)$.
The cross section is therefore differential in  $\bm{P}$ and $\bm{k}$.
The sum over the spins of $D^{*0}$ and $\bar D^{*0}$  is  implicit.
We assume the $D^{*0} \bar D^{*0}$ channel has no resonant threshold enhancement 
analogous to that from the  $X(3872)$ in the $D^{*0} \bar D^{0}$ channel, 
so $\mathcal{A}^{ij}_{D^{*0} \bar D^{*0} +y}$ can be interpreted as a short-distance amplitude. 

In the prompt production of a pair of charm mesons with large transverse momentum at a high-energy hadron collider,
isospin symmetry and heavy-quark spin  symmetry imply that the short-distance
amplitudes for producing each pair of spin states for the charm mesons $D^{(*)} \bar D^{(*)}$ should be equal.
Since we have absorbed  factors of 3 from summing over spin states of $D^*$ and $\bar D^*$
into the amplitudes, the squared  amplitude $| \mathcal{A}_{D^{*0} \bar D^{*0} +y}|^2$  
in Eq.~\eqref{factor-DstarDstar} should  be equal to 3 times the 
 corresponding squared short-distance amplitude  $|\mathcal{A}_{D^{*0} \bar D^{0} +y}^{\mathrm{s.d.}}|^2$
in the expression for the $X$ cross section in Eq.~\eqref{factor-X-interference}.
For inclusive prompt production at the Tevatron or LHC,
 the total momentum of $D^{*0} \bar D^{*0}$ is large enough that the difference 
between the mass $2 M_{*0}$ of   $(D^*\bar D^*)$
and the mass $M_{*0}+M_0$ of  $(D^*\bar D)$ can be neglected in the phase space integrals.
We can therefore eliminate the short-distance factor in Eq.~\eqref{factor-DstarDstar}
in favor of the $X$ cross section in Eq.~\eqref{factor-X-interference}
at  the expense of introducing the unknown momentum scale $\Lambda$ 
of order $m_\pi$ and the unknown binding momentum $\gamma_X$:
%===============
\begin{equation}
d\sigma[D^{*0} \bar D^{*0}]  \approx
d\sigma[X(3872)] \, \frac{12\pi\mu}{\gamma_X \Lambda^2}\, 
 \frac{d^3k}{(2 \pi)^3 M_{*0}}.
\label{sigmaDstarDstar-X}
\end{equation}
%===============

%%%%%%%%%%%%%%%%%%%%%%%%%%%%%%%%%%%%%%%%%%%%%%%%
\begin{figure}[t]
\includegraphics*[width=0.4\linewidth]{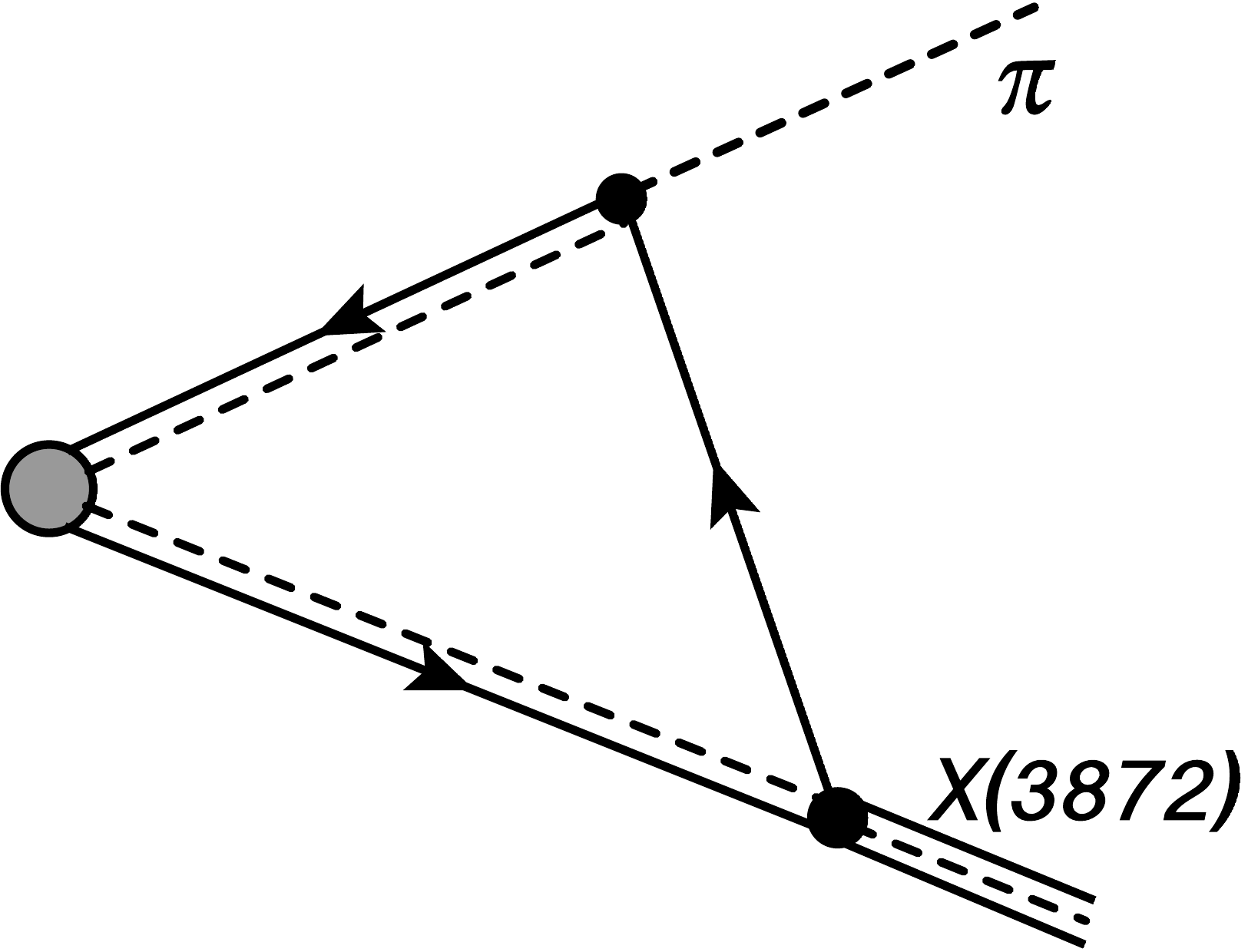} ~
\includegraphics*[width=0.4\linewidth]{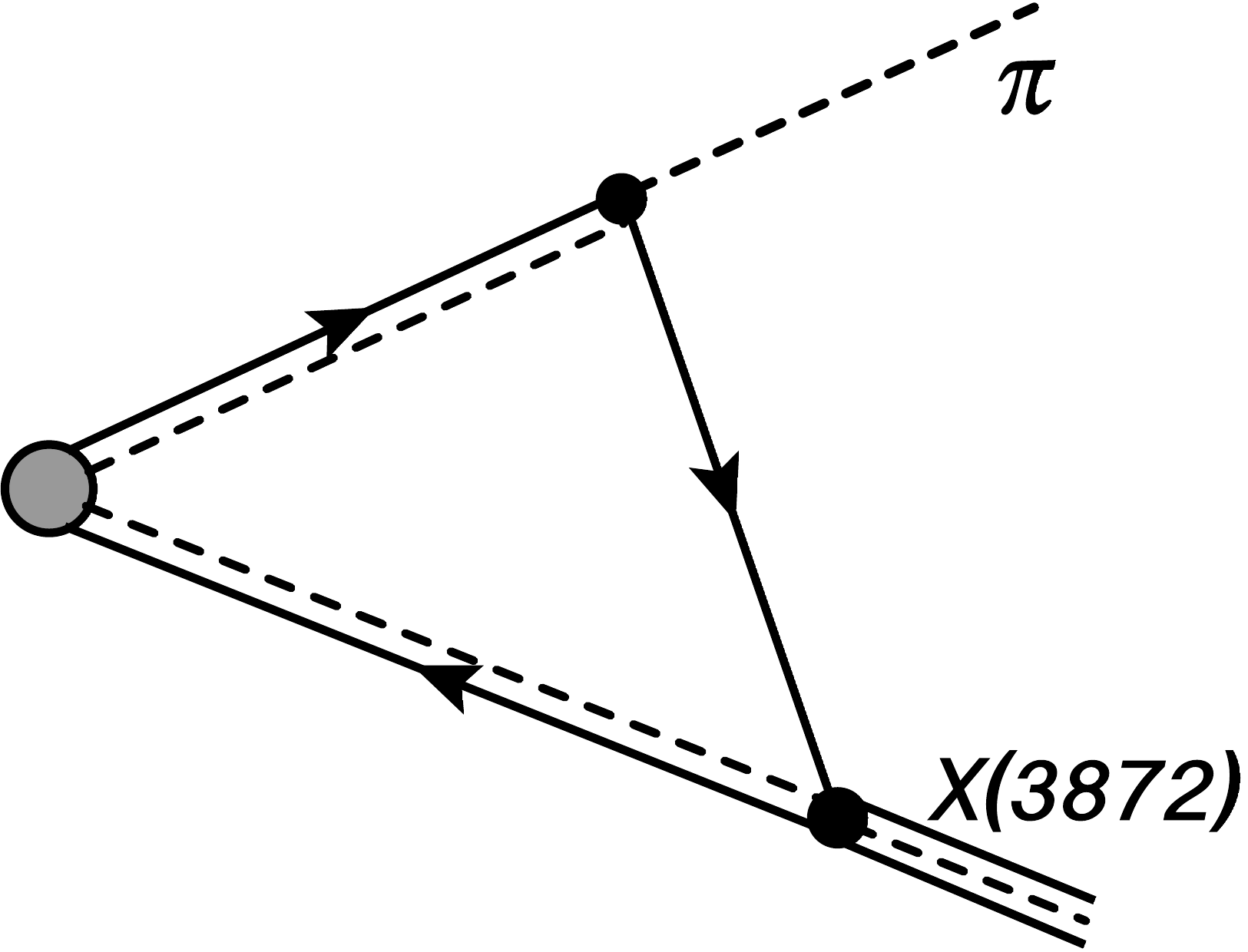} 
\caption{
Feynman diagrams in XEFT for $D^{*0} \bar D^{*0}$ created at a point to rescatter into $X\pi^0$.
The $X(3872)$ is represented by a triple line with two solid lines and one dashed line.
The $D$ and $\bar D$ are represented by solid lines.
The $\pi$ is represented by a dashed line.
}
\label{fig:DDtoXpi}
%\vspace*{0.0cm}
\end{figure}
%%%%%%%%%%%%%%%%%%%%%%%%%%%%%%%%%%%%%%%%%%%%%%

%\newpage

%%%%%%%%%%%%%%%%%%%%%%%%%%%%%%%%%%%%%%%%%%
\section{Production of  $\bm{X\pi}$ near $\bm{D^* \bar{D}^*}$ Threshold}
\label{sec:Xsoftpi}
%%%%%%%%%%%%%%%%%%%%%%%%%%%%%%%%%%%%%%%%%%

A pair of spin-1 charm mesons $D^{*0} \bar D^{*0}$ created at short distances can rescatter 
into $X(3872)\,\pi^0$. The Feynman diagrams for $D^{*0} \bar D^{*0}$ created at a point to rescatter 
into $X\pi^0$ are shown in Fig.~\ref{fig:DDtoXpi}.
The amplitude for these diagrams can be calculated using XEFT.
In addition to the Feynman rules for Galilean-invariant XEFT  given in \cite{Braaten:2015tga},
we need the vertices for the coupling of $D^{*0} \bar D^0$ and $D^{0} \bar D^{*0}$ to $X$.
The vertices are given by $(\sqrt{\pi \gamma_X}/\mu) \delta^{ij}$, where $i$ and $j$ are the spin indices of the  
spin-1 charm meson and the $X$ \cite{Braaten:2010mg}.
If the amplitude for producing  $D^{*0} \bar D^{*0}$ at short distances with specific polarization vectors 
is $\mathcal{A}^{ij}_{D^{*0} \bar D^{*0}+y} \varepsilon^i \bar \varepsilon^j$,
the vertex for creating $D^{*0} \bar D^{*0}$ at a point in Fig.~\ref{fig:DDtoXpi} 
is $i \mathcal{A}^{ij}_{D^{*0} \bar D^{*0}+y}$.
That amplitude can be treated as a point vertex, because the momenta of the colliding hadrons 
and of the additional particles $y$ in the $X\pi$ rest frame are all large compared to 
the relative momentum $\bm{q}$ of $X$ and $\pi$.
The integral over the loop energy is conveniently evaluated by contours using the pole of the 
propagator for the $D^{*0}$ or $\bar D^{*0}$ line attached to the $X$.
The resulting expression for the amplitude 
for producing $X\pi^0$ with small relative momentum $\bm{q}$  in the $X \pi$ rest frame 
and with polarization vector $\bm{\varepsilon}$  for  the $X$ is
%===============
\begin{eqnarray}
&&i\, \mathcal{A}^{ij}_{D^{*0} \bar D^{*0}+y}
 \frac{g (\pi \gamma_X /m_0)^{1/2} }{2  \mu f_\pi}
  \int \frac{d^3k}{(2\pi)^3} \
 \left[  \frac{(M_0+m_0) q^i - m_0 k^i}{M_0+m_0} \varepsilon^j  + \varepsilon^i  \frac{(M_0+m_0)q^j - m_0 k^j}{M_0+m_0} \right]
\nonumber\\
 && \hspace{5cm}
 \times
  \frac{1}{-\gamma_X^2/(2\mu) + i \Gamma_{*0}/2 
  - \mu \big( (\bm q - \bm k)/M_0 - \bm k/(M_0+m_0) \big)^2/2}
\nonumber\\
 && \hspace{5cm}
 \times
 \frac{1}{-\gamma_X^2/(2\mu) - \delta_0 + i \Gamma_{*0} 
 + \bm q^2/(2\mu_{X\pi})  - \bm k^2/(M_0+m_0)},
\label{amplitudeXpi0intk}
\end{eqnarray}
%===============
where $\mu = M_0(M_0+m_0)/(2M_0+m_0)$ and $\mu_{X \pi} = (2M_0+m_0)m_0/(2(M_0+m_0))$ 
are the Galilean-invariant reduced masses of $D^{*0}\bar D^0$ and $X\pi$.
The integral is over the loop momentum $\bm k$
of the $D^{*0}$ or $\bar D^{*0}$ that becomes a constituent of the $X$.
The remaining two propagators can be combined into a single denominator by introducing 
an integral over a Feynman parameter.  
The integral over the loop momentum can  be evaluated analytically.  The resulting amplitude
for producing $X\pi^0$  is
%===============
\begin{eqnarray}
&&i\, \mathcal{A}^{ij}_{D^{*0} \bar D^{*0}+y}
 \frac{g (\pi \gamma_X /m_0)^{1/2} M_{*0}^{3/2}}{16 \pi  \mu f_\pi}
 \big(\varepsilon^i q^j + q^i \varepsilon^j \big)
 \int_0^1 dx \left( \frac{2M_0}{2M_0+(1-x)m_0} \right)^{5/2}
\nonumber\\
 && \times
 \Bigg[ \big(\delta_0-\gamma_X^2/2\mu\big) - (1+x)\big(\delta_0-i \Gamma_{*0}/2\big)
+\frac{xM_0}{(2M_0+(1-x)m_0)\mu_{X\pi}}  \bm{q}^2 \Bigg]^{-1/2},~~~~~~
\label{amplitudeXpi0int}
\end{eqnarray}
%===============
where $\delta_0=M_{*0} - M_0- m_0 = 7.0$~MeV,
$\Gamma_{*0} \approx 60$~keV is the predicted decay width of $D^{*0}$ \cite{Braaten:2015tga},
and $g/(2\sqrt{m_0} f_\pi)= 0.30/m_0^{3/2}$  is the coupling constant for the pion-emission vertex  \cite{Braaten:2015tga}.
The final integral over $x$ can also be evaluated analytically
if the integrand is simplified using $m_0 \ll M_0$.  
This reduces the prefactor of $\bm{q}^2$ in the denominator of the integrand to $x/2m_0$.
Our final result for the amplitude  is rather simple:
%===============
\begin{equation}
i\, \mathcal{A}^{ij}_{D^{*0} \bar D^{*0}+y}
 \frac{g (\pi \gamma_X M_{*0}^3/m_0)^{1/2}}{8 \pi  \mu f_\pi} 
\frac{\varepsilon^i q^j + q^i \varepsilon^j }
{\sqrt{q^2/2m_0  - \delta_0 -\gamma_X^2/2\mu + i \Gamma_{*0}} + \sqrt{-\gamma_X^2/2\mu + i \Gamma_{*0}/2}}.
\label{amplitudeXpi0}
\end{equation}
%===============
The denominator is a kinematic singularity factor that would have a zero at  $q^2= 2 m_0 \delta_0$
if the binding momentum $\gamma_X$ and the width $\Gamma_{*0}$ were both zero.
The kinematic singularity is called a {\it triangle singularity},
because it arises from the region where the three charm-meson lines that form a triangle 
in the Feynman diagrams in Fig.~\ref{fig:DDtoXpi} are all simultaneously on shell.

A pair of spin-1 charm mesons $D^{*+} \bar D^{*0}$ created at short distances can rescatter 
into $X\pi^+$.  The Feynman diagram is  the second diagram in Fig.~\ref{fig:DDtoXpi},
with the virtual $D^{*0}$ line replaced by a $D^{*+}$ line  and the final-state $\pi^0$ replaced by $\pi^+$.
The integral over the loop momentum can  be evaluated analytically.
The final integral over $x$ can also be evaluated analytically
if the integrand is simplified using $m_0 \ll M_0$.  
Our final result for the amplitude  is rather simple:
%===============
\begin{eqnarray}
&&i\, \mathcal{A}^{ij}_{D^{*+} \bar D^{*0}+y}
 \frac{g (2\pi \gamma_X M_{*0}^3/m_0)^{1/2}}{8 \pi  \mu f_\pi} 
 \nonumber\\
 && \hspace{2cm} \times
  \frac{q^i \varepsilon^j}{\sqrt{q^2/2m_0  - \delta_1  -\gamma_X^2/2\mu+ i (\Gamma_{*0}+ \Gamma_{*1})/2}  +
\sqrt{ -\gamma_X^2/2\mu + i \Gamma_{*0}/2}},
\label{amplitudeXpi+}
\end{eqnarray}
%===============
where $\delta_1=M_{*1} - M_0-m_1= 5.9$~MeV
and $\Gamma_{*1} \approx 83$~keV is the measured decay width of $D^{*+}$.
The pair of spin-1 charm mesons $D^{*0} D^{*-}$ created at short distances can rescatter into $X\pi^-$.  
The amplitude is the same as that in Eq.~\eqref{amplitudeXpi+} except that the
short-distance factor is replaced by  $\mathcal{A}^{ij}_{D^{*0} D^{*-}+y}$
and $q^i \varepsilon^j$ is replaced by $\varepsilon^i q^j$. 
The denominator of Eq.~\eqref{amplitudeXpi+} is a triangle-singularity factor 
that would have a zero at  $q^2= 2 m_0 \delta_1$
if the binding momentum $\gamma_X$ and the widths $\Gamma_{*0}$ and $\Gamma_{*1}$ were all zero.

%%%%%%%%%%%%%%%%%%%%%%%%%%%%%%%%%%%%%%%%%%%%%%%%
\begin{figure}[t]
\includegraphics*[width=0.8\linewidth]{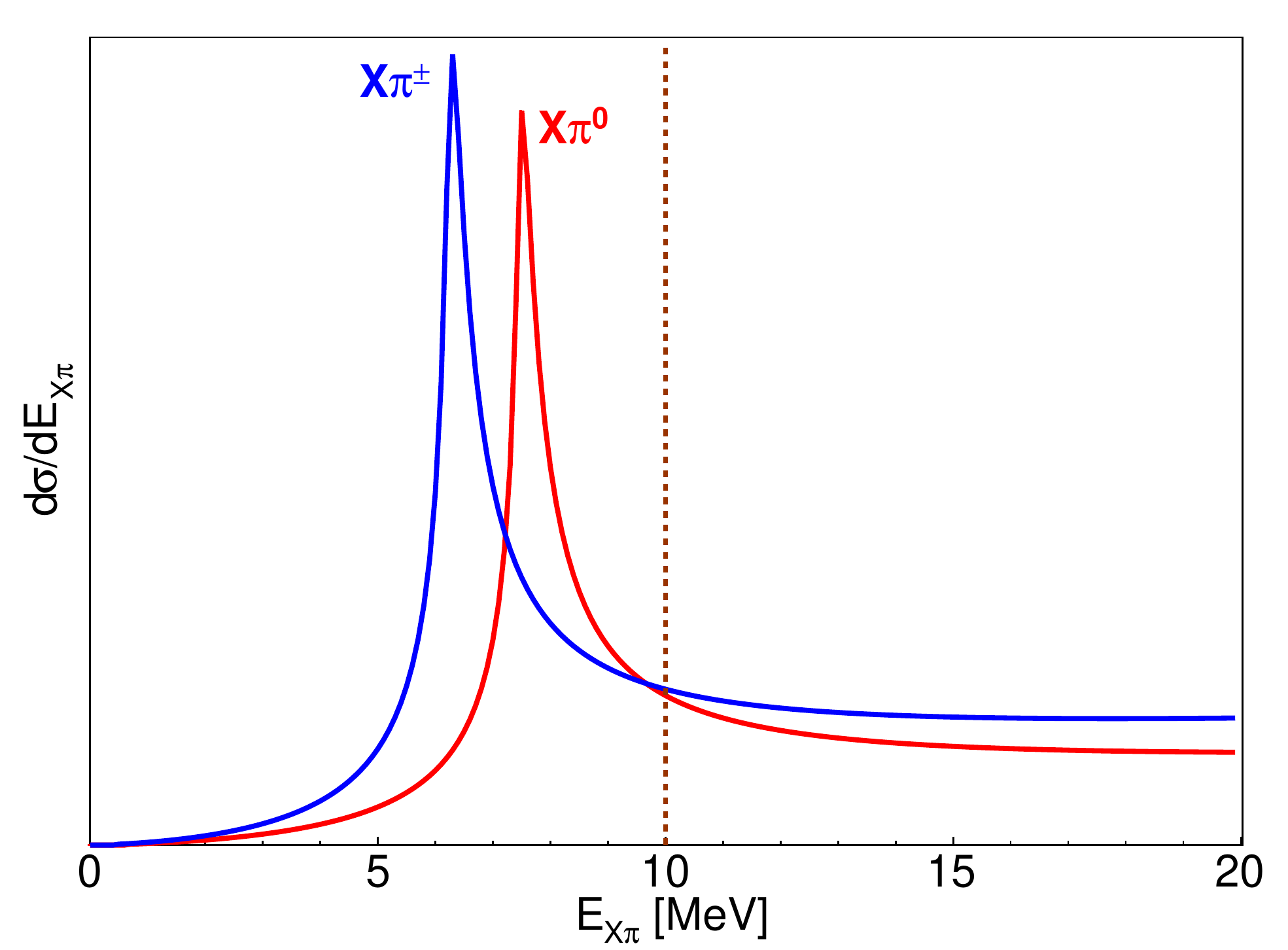} 
\caption{
Differential cross sections $d\sigma/dE_{X\pi}$ as  functions of the kinetic energy $E_{X\pi}$
for  $X\pi^0$ and for $X\pi^\pm$ summed over $\pm$ 
from rescattering of  $D^* \bar D^*$ created at short distances.
The binding energy of the $X$ is 0.17~MeV.
The region of validity of XEFT extends out to about the vertical dotted line
at $E_{X\pi} = 10$~MeV.
The scale on the vertical axis is arbitrary,
but the relative normalizations of the two curves is determined.
}
\label{fig:Xpi-q}
%\vspace*{0.0cm}
\end{figure}
%%%%%%%%%%%%%%%%%%%%%%%%%%%%%%%%%%%%%%%%%%%%%%

Our expressions for the amplitudes  in Eqs.~\eqref{amplitudeXpi0} and \eqref{amplitudeXpi+}
should be accurate provided the  momentum integral 
in Eq.~\eqref{amplitudeXpi0intk}
is dominated by regions where the relative momentum $\bm{k}$ of the charm mesons that form the $X$ 
is less than about $m_\pi$. 
This condition imposes a constraint on the relative momentum $\bm q$ of $X$ and $\pi$.
The  constraint can be deduced from the integrated expression in Eq.~\eqref{amplitudeXpi0int}  
by requiring the energy proportional to $\bm q^2$ inside the last factor to be less than $m_\pi^2/2\mu$.
At $x=1$, this energy is $E_{X\pi} = q^2/2\mu_{X\pi}$. 
Thus the kinetic energy $E_{X\pi}$ must be less than about $m_\pi^2/2\mu \approx 10$~MeV.

To obtain the differential cross section for producing $X\pi^0$ and $X\pi^+$,
the amplitudes in Eqs.~\eqref{amplitudeXpi0} and \eqref{amplitudeXpi+} must be multiplied by their complex conjugates.  
Their product must be summed over the additional particles $y$
(which includes sums over their spin states and integrals over their momenta),
summed over  the spin states of $X$,
integrated over the phase space of $X$ and $\pi$, and divided by the flux factor.
For each set of additional particles $y$,
the product of $\mathcal{A}^{ij}_{D^{*} \bar D^{*}+y}$ and $(\mathcal{A}^{kl}_{D^{*} \bar D^{*}+y})^*$
summed over their spins  and integrated over their momenta, is a density matrix in the spin indices $ij$ and $kl$.
In the case of inclusive prompt production at the Tevatron or the LHC,  cancellations from the sum 
over the many additional particles $y$ leaves a density matrix diagonal in the indices $i k$ and in the indices $jl$, 
 as in Eq.~\eqref{AASS-D*Dbar*}.
After multiplying by the  tensors in Eqs.~\eqref{amplitudeXpi0} and \eqref{amplitudeXpi+} that involve polarization vectors,
the sum over the spin states of $X$ results in a factor $\bm{q}^2$:
%===============
\begin{subequations}
\begin{eqnarray}
&&\sum_\mathrm{spins} \left\langle\mathcal{A}^{ij}_{D^{*0} \bar D^{*0}+y} 
\big(\mathcal{A}^{kl}_{D^{*0} \bar D^{*0}+y}\big)^* \right\rangle
(\varepsilon^i q^j + q^i \varepsilon^j ) (\varepsilon^k q^l + q^k \varepsilon^l )^*
=   \frac89  \left\langle\big|  \mathcal{A}_{D^{*0} \bar D^{*0}+y}\big|^2 \right\rangle \bm{q}^2,
\label{sumAASS-D*Dbar*0}
\\
&&\sum_\mathrm{spins}\left\langle\mathcal{A}^{ij}_{D^{*+} \bar D^{*0}+y} 
\big(\mathcal{A}^{kl}_{D^{*+} \bar D^{*0}+y}\big)^*\right\rangle
(q^i \varepsilon^j ) (q^k \varepsilon^l )^*
=   \frac13 \left\langle  \big|  \mathcal{A}_{D^{*+} \bar D^{*0}+y}\big|^2  \right\rangle \bm{q}^2.
\label{sumAASS-D*Dbar*pm}
\end{eqnarray}
\label{sumAASS}%
\end{subequations}
%===============
The  squared amplitude $|  \mathcal{A}_{D^{*0} \bar D^{*0}+y}|^2$ in Eq.~\eqref{sumAASS-D*Dbar*0}
also appears in the short-distance factor in the cross section for producing  $D^{*0} \bar D^{*0}$
in Eq.~\eqref{factor-DstarDstar}.  
The short-distance factor in the cross section for producing $X \pi^0$
can therefore be eliminated in favor of the differential cross section for producing  $D^{*0} \bar D^{*0}$
near threshold given by Eq.~\eqref{factor-DstarDstar}.
Alternatively the short-distance factor  can be eliminated in favor of the cross section for producing $X$ by using Eq.~\eqref{sigmaDstarDstar-X}.

The differential cross sections for producing $X\pi^0$ and $X \pi^+$
with small relative momentum $\bm{q}$ can be expressed as
%===============
\begin{subequations}
\begin{eqnarray}
\frac{d\sigma}{d^3q}[X \pi^0] &=& \sigma[X]
\frac{g^2 M_{*0}^3}{48\pi^3 \mu\mu_{X\pi} f_\pi^2 \Lambda^2} 
\nonumber\\
 &&\hspace{0cm}
 \times\frac{q^2/2m_0}
{\big| \sqrt{q^2/2m_0  - \delta_0-\gamma_X^2/2\mu  + i \Gamma_{*0}} + \sqrt{-\gamma_X^2/2\mu+ i \Gamma_{*0}/2} \,\big|^2},
\label{dsigmaXpi0}
\\
\frac{d\sigma}{d^3q}[X  \pi^+] &=&\sigma[X]
\frac{g^2 M_{*0}^3}{64 \pi^3 \mu\mu_{X\pi}  f_\pi^2 \Lambda^2} 
\nonumber \\
&&\hspace{0cm}
 \times\frac{q^2/2m_0}
{\big| \sqrt{q^2/2m_0  - \delta_1  -\gamma_X^2/2\mu+ i (\Gamma_{*0}+ \Gamma_{*1})/2}  +
\sqrt{ -\gamma_X^2/2\mu + i \Gamma_{*0}/2} \big|^2}.
\label{dsigmaXpi+}
\end{eqnarray}
\label{dsigmaXpi}%
\end{subequations}
%===============
We have used Eq.~\eqref{sigmaDstarDstar-X} to
eliminate the short-distance factors  in favor of the cross section for producing $X$
at  the expense of introducing the unknown momentum scale $\Lambda$  of order $m_\pi$.
The factors of the unknown binding momentum $\gamma_X$ have canceled.
The differential  cross section for producing $X\pi^-$ is also given by Eq.~\eqref{dsigmaXpi+}.

A convenient kinematic variable to consider is the kinetic energy $E_{X\pi} = q^2/2 \mu_{X\pi}$ 
of  $X\pi$ in the $X\pi$ CM frame, since
the square of the $X\pi$ invariant mass near the $X\pi$ threshold  is approximately linear in $E_{X\pi}$.
The dependence of the differential cross sections $d\sigma/dE_{X\pi}$ on  $E_{X\pi}$
is illustrated in Fig.~\ref{fig:Xpi-q} for $\gamma_X = 18$~MeV.
The differential cross section for $X\pi^0$ in Fig.~\ref{fig:Xpi-q} has a  narrow peak in $E_{X\pi}$ 
near $\delta_0 = 7.0$~MeV that  comes from a  charm-meson triangle singularity.
The full width at half maximum of the denominator factor in Eq.~\eqref{dsigmaXpi0} 
is approximately $1.17 \, \gamma_X^2/2\mu$
if the binding energy $\gamma_X^2/2\mu$  is large compared to $\Gamma_{*0}$
and approximately $6.21 \, \Gamma_{*0} \approx 370$~keV if $\gamma_X^2/2\mu$  is small compared to $\Gamma_{*0}$.
The differential cross section for $X\pi^+$  in Fig.~\ref{fig:Xpi-q} has  a  narrow peak in  
$E_{X\pi}$ near $\delta_1 = 5.9$~MeV that comes from a   charm-meson triangle singularity.
The full width at half maximum of the denominator factor in Eq.~\eqref{dsigmaXpi+} 
is approximately $1.17 \, \gamma_X^2/2\mu$
if $\gamma_X^2/2\mu$  is large compared to $\Gamma_{*0}$  and $\Gamma_{*1}$ and approximately 430~keV
if $\gamma_X^2/2\mu$  is small compared to $\Gamma_{*0}$ and $\Gamma_{*1}$.

%%%%%%%%%%%%%%%%%%%%%%%%%%%%%%%%%%%%%%%%%%%%%%%%
\begin{figure}[t]
\includegraphics*[width=0.8\linewidth]{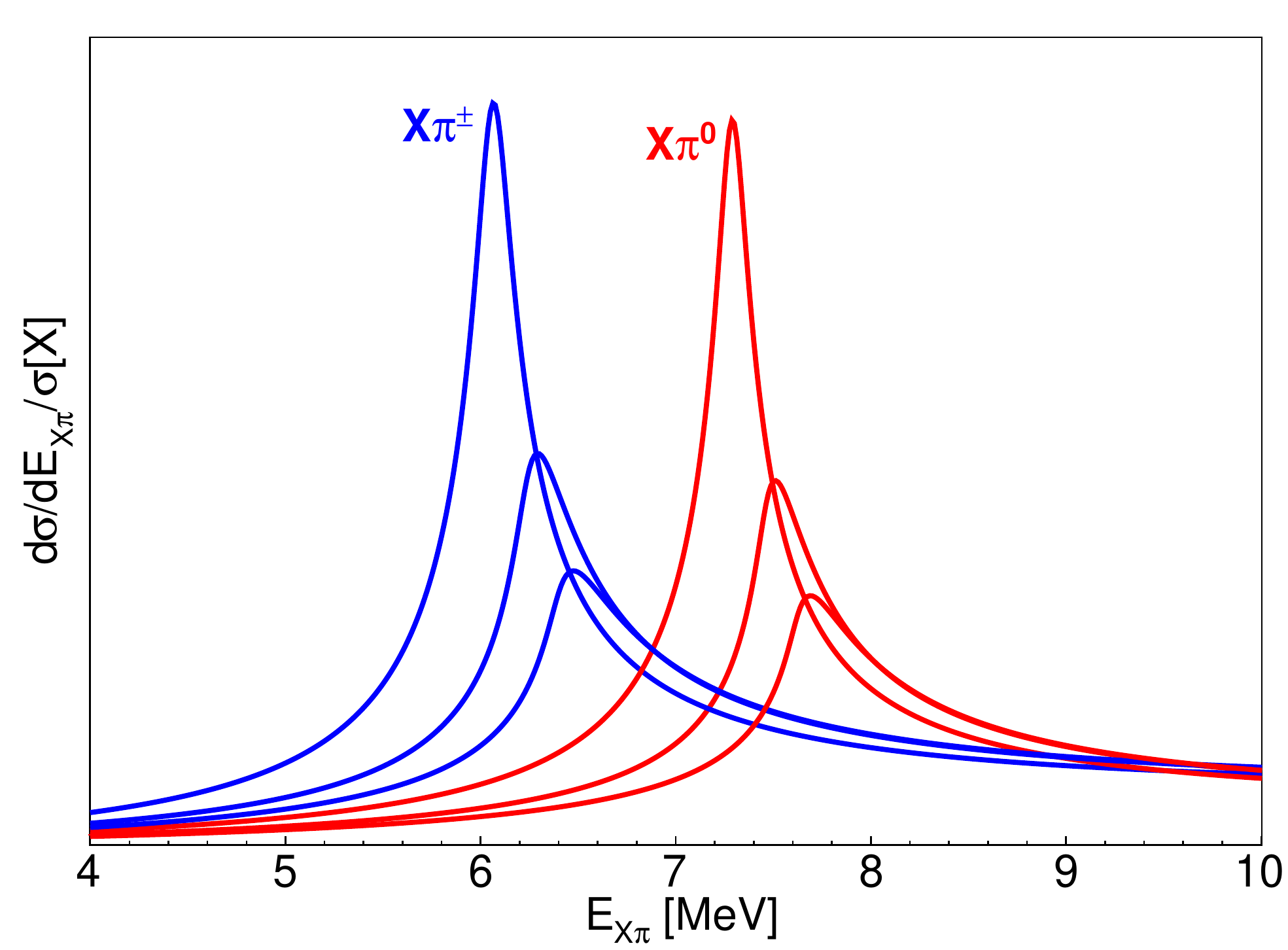} 
\caption{
Differential cross sections $d\sigma/dE_{X\pi}/\sigma[X]$ as functions of 
the  kinetic energy $E_{X\pi}$ for $X\pi^0$  (right curves) and for $X\pi^\pm$  summed over $\pm$ (left curves) .
In order of decreasing height, the curves are for binding energies 
$E_X=0$, $|E_X|=0.17$~MeV, and $|E_X|=0.34$~MeV. 
The scale on the vertical axis is arbitrary,
but the relative normalizations of the curves are determined.
}
\label{fig:Xpi-EX}
%\vspace*{0.0cm}
\end{figure}
%%%%%%%%%%%%%%%%%%%%%%%%%%%%%%%%%%%%%%%%%%%%%%

The sensitivity of the narrow peaks in Fig.~\ref{fig:Xpi-q} to 
the binding energy $|E_X| = \gamma_X^2/2\mu$ is illustrated in Fig.~\ref{fig:Xpi-EX}.
The differential cross section $d\sigma/dE_{X\pi}$ has a factor of the square root of $|E_X|$
that  is removed by dividing  by $\sigma[X]$. 
If $|E_X|$ is decreased from 0.17~MeV to 0,  the height of the peak increases by about a factor of 2.
If $|E_X|$ is increased from 0.17~MeV to 0.34~MeV, the height of the peak decreases by about a factor of 0.7.
The areas under the peaks are less sensitive to the binding energy. 

The contributions of the triangle singularities to the integrated cross sections for $X\pi$ 
can be estimated by integrating the momentum distributions in Eqs.~\eqref{dsigmaXpi}
from the threshold to some energy  $E_\mathrm{max}$ beyond the peak.
In the limits $\Gamma_{*0} \to 0$ and $\gamma_X \ll \sqrt{\mu \delta_0}$,
the integral of the momentum dependent factor in Eq.~\eqref{dsigmaXpi0}
over the region   $|\bm q| <  q_\mathrm{max}$ is
%===============
\begin{eqnarray}
&&\int_{q < q_\mathrm{max}}  \frac{d^3q}{(2\pi)^3} \frac{q^2/2m_0}
{\big| \sqrt{q^2/2m_0  - \delta_0 -\gamma_X^2/2\mu + i \epsilon} + i \sqrt{\gamma_X^2/2\mu}  \big|^2}
= \frac{1}{2 \pi^2} (2m_0 \delta_0)^{3/2} 
\nonumber\\
&& \hspace{0.5cm}
\times \left[  \log \frac{8 \mu\delta_0}{\gamma_X^2} + \frac13 \left( \frac{q_\mathrm{max}^2}{2m_0\delta_0} \right)^{3/2} 
+\left( \frac{q_\mathrm{max}^2}{2m_0\delta_0} \right)^{1/2}
- \frac12 \log \frac{\sqrt{q_\mathrm{max}^2/2m_0\delta_0} +1}{\sqrt{q_\mathrm{max}^2/2m_0\delta_0}-1}  -  \frac{11}{3} \right].
\label{intsoftpi}
\end{eqnarray}
%===============
The coefficient of $(2m_0 \delta_0)^{3/2}$ diverges logarithmically as $\gamma_X \to 0$.
If we do not take the limit $\Gamma_{*0} \to 0$,
the coefficient of $(m_0 \delta_0)^{3/2} $ also depends logarithmically on $\Gamma_{*0}$.
In the limit $\Gamma_{*0} \to 0$ and $\Gamma_{*1} \to 0$,
the integral of the momentum dependent factor in Eq.~\eqref{dsigmaXpi+} is
given by  Eq.~\eqref{intsoftpi} with $\delta_0$ replaced by $\delta_1$.

We can use our expressions for the differential cross sections for $X \pi$  in Eq.~\eqref{dsigmaXpi}
to estimate the cross sections for producing $X\pi$ in the peak from the triangle singularity.
We denote the region of the peak by $(X\pi)_\triangle$.
We declare that region to be $E_{X\pi}$ from 0 up to $E_\mathrm{max}= 2 \delta_0$ for  $(X\pi^0)_\triangle$
and up to $E_\mathrm{max}= 2 \delta_1$ for  $(X\pi^\pm)_\triangle$.
We approximate the integrals over the momentum distributions in 
Eqs.~\eqref{dsigmaXpi0} and \eqref{dsigmaXpi+} using the integral in 
Eq.~\eqref{intsoftpi} and the analogous integral with $\delta_0$ replaced by $\delta_1$.
The resulting estimates of the ratios of the cross sections for $(X \pi)_\triangle$ to the cross section for $X$
as a function of the binding energy $|E_X|=\gamma_X^2/2\mu$ is
%===============
\begin{subequations}
\begin{eqnarray}
\frac{\sigma\big[  (X \pi^0)_\triangle \big]}{\sigma[X]}  \approx
0.049 \left( \frac{m_\pi}{\Lambda} \right)^2
\left[ 2.82 - \log \frac{|E_X|}{0.17~\mathrm{MeV}} \right] ,
\label{sigmaXpi0}
\\
\frac{\sigma\big[  (X \pi^+)_\triangle \big]}{\sigma[X]}  \approx
0.028 \left( \frac{m_\pi}{\Lambda} \right)^2
\left[ 2.64 - \log \frac{|E_X|}{0.17~\mathrm{MeV}} \right] .
\label{sigmaXpi+}
\end{eqnarray}
\label{sigmaXpi0/+}%
\end{subequations}
%===============
The prefactors were obtained by setting $m_\pi = m_0$.
The cross section for producing $(X \pi^-)_\triangle$ should be the same  as that for producing $(X \pi^+)_\triangle$.
If the binding energy of $X$ is not too far from 0.17~MeV,
the ratio of the cross section for producing $(X \pi^\pm)_\triangle$ summed over $\pm$
to the cross section for producing $X$ without an accompanying pion is predicted to be  $0.14\, (m_\pi/\Lambda)^2$. 
It is only logarithmically sensitive to the binding energy $|E_X|$,
but it depends strongly on the unknown momentum scale $\Lambda$ of order $m_\pi$.

The inclusive prompt production of $X$ in $p p$ collisions has been studied at the 
Large Hadron Collider (LHC) by the CMS \cite{Chatrchyan:2013cld}
and ATLAS \cite{Aaboud:2016vzw} collaborations.
The inclusive production of $X$ in $p p$ collisions has also been studied at the LHC by the LHCb \cite{Aaij:2011sn}, 
but they did not separate the prompt production from production from decay of bottom hadrons.
The CMS collaboration measured the $\pi^+\pi^-$ invariant mass distribution from the decays 
of $X$  into $J/\psi\, \pi^+ \pi^-$ \cite{Chatrchyan:2013cld}.
They reported the $X$ yield in the appropriate kinematic region to be $6302 \pm 346$ events.
If the binding energy of $X$ is 0.17~MeV, the yield in $(X \pi^\pm)_\triangle$ summed over $\pm$
 is predicted by Eq.~\eqref{sigmaXpi+} to be smaller by the factor $0.14 (m_\pi/\Lambda)^2$. 
 Despite the uncertainty from the unknown scale $\Lambda$,
this is large enough to encourage a search for the peak from the charm-meson triangle singularity at the LHC. 

%\newpage

%%%%%%%%%%%%%%%%%%%%%%%%%%%%%%%%%%%%%%%%%%
\section{Summary and Discussion}
\label{sec:Summary}
%%%%%%%%%%%%%%%%%%%%%%%%%%%%%%%%%%%%%%%%%%

We have discussed the inclusive prompt production of the $X(3872)$ at high energy hadron colliders under the assumption 
that the $X$ is a weakly bound charm-meson molecule with the particle content in Eq.~\eqref{Xflavor}.
We considered the production of $X$ through the creation of a charm-meson pair at short distances
of order $1/m_\pi$ or smaller. The formation of the $X$ proceeds on longer distance scales, 
and it can be described by the effective field theory XEFT.
The $X$ can be produced by the creation of its constituents $D^{*0} \bar D^0$ and $D^0 \bar D^{*0}$ at short distances
followed by the binding of the charm mesons into $X$.
The $X$ can also be produced by the creation of a pair  of spin-1 charm mesons $D^* \bar D^*$ at short distances
followed by the rescattering of the charm mesons into $X\pi$.

The differential cross sections for the inclusive prompt production of $X\pi^0$ from rescattering of $D^{*0}\bar D^{*0}$
and of $X\pi^+$ from rescattering of $D^{*+}\bar D^{*0}$ are given in Eqs.~\eqref{dsigmaXpi}
and illustrated in Figure~\ref{fig:Xpi-EX}.
They have a narrow peak from a  charm-meson triangle singularity at a kinetic energy $E_{X\pi}$ 
close to the $D^{*0}\bar D^{*0}$ and $D^{*+}\bar D^{*0}$ scattering thresholds, respectively.
Estimates of the ratio of the cross sections for producing  $X\pi^0$ and $X\pi^\pm$ 
in the peak from the charm-meson triangle singularity
to the cross section for producing $X$ without an accompanying pion are given in Eqs.~\eqref{sigmaXpi0/+}.

The prompt production of $X$ accompanied by $\pi^+$ or $\pi^-$ can be studied at a hadron collider.
The charged pion provides a clean signature for this new production mechanism.
For prompt production at the LHC, the observation of this signature may be complicated by the 
combinatorial background from the hundreds of pions produced at the primary interaction vertex.
The combinatorial background is especially severe in forward production, which is measured by the LHCb detector.
For central production at large transverse momentum, which is measured by the ATLAS and CMS  detectors,
the combinatorial background may be more manageable.
Given the number of $X$ events that have been observed,
the number of events in the peak in the $X\pi^\pm$  invariant mass distribution 
from the charm-meson triangle singularity can be estimated using Eq.~\eqref{sigmaXpi+}.
The estimate is large enough to encourage the effort to observe this peak at the LHC.
The observation of
such a  peak  would provide strong support for the identification 
of $X$ as  a weakly bound charm-meson molecule and present a serious challenge to other models.

%%%%%%%%%%%%%%%%%%%%%%%%%%%%%%%%%%%%%%%%%%
\begin{acknowledgments}
This work was supported in part by the Department of Energy under grant DE-SC0011726 and by the National Science Foundation under grant PHY-1607190.
We thank T.~Skwarnicki for valuable comments.
\end{acknowledgments}
%%%%%%%%%%%%%%%%%%%%%%%%%%%%%%%%%%%%%%%%%%

%\end{linenumbers}

%\newpage

%%%%%%%%%%%%%%%%%%%%%%%%%%%%%%%%%%%%%%%%%%

%%%%%%%%%%%%%%%%%%%%%%%%%%%%%%%%%%%%%%%%%%

\end{document}